\documentclass[prl,aps]{revtex4}
\usepackage{psfrag}
\usepackage{graphicx}
\usepackage{dcolumn}
\usepackage{color}
\usepackage{latexsym,amsfonts}
\usepackage{bm}
\usepackage{amssymb}
\begin{document}

\title{Energy Level Displacement\\ of Excited $np$ State of Kaonic
Deuterium\\ In Faddeev Equation Approach}

\author{M. Faber${^a}$, M. P. Faifman${^b}$, A. N. Ivanov${^a}$,
  J. Marton${^c}$, M. Pitschmann${^{a,d,e}}$, N. I. Troitskaya${^f}$}
\affiliation{${^a}$Atominstitut, Technische Universit\"at Wien,
  Wiedner Hauptstrasse 8-10, A-1040 Wien, Austria}
\affiliation{${^b}$National Research Centre ``Kurchatov Institute'',
  Moscow 123182, Russia Federation} \affiliation{${^c}$Stefan Meyer
  Institut f\"ur subatomare Physik \"Osterreichische Akademie der
  Wissenschaften, Boltzmanngasse 3, A-1090, Wien,
  Austria}\affiliation{$^d$University of Wisconsin--Madison,
  Department of Physics, 1150 University Avenue, Madison, WI 53706,
  USA}\affiliation{$^c$Physics Division, Argonne National Laboratory,
  Argonne, Illinois 60439, USA} \affiliation{ $^f$State Polytechnic
  University of St. Petersburg, Polytechnicheskaya 29, 195251, Russian
  Federation} \email{ivanov@kph.tuwien.ac.at}

\date{\today}

  \begin{abstract}
 We calculate the energy level displacement of the excited $np$ state
 of kaonic deuterium in terms of the P--wave scattering length of
 $K^-d$ scattering.  We solve the Faddeev equations for the amplitude
 of $K^-d$ scattering in the fixed centre approximation and derive the
 complex P--wave scattering length of $K^-d$ scattering in terms of
 the S--wave and P--wave scattering lengths of $\bar{K}N$
 scattering. The estimated uncertainty of the complex P--wave
 scattering length is of about $15\,\%$.  For the calculated width
 $\Gamma_{2p} = 10.203\,{\rm meV}$ of the excited $2p$ state of kaonic
 deuterium we evaluate the yield $Y_{K^-d} = 0.27\,\%$ of $X$--rays
 for the $K_{\alpha}$ emission line of kaonic deuterium.  Using the
 complex S--wave and P--wave scattering lengths of $\bar{K}N$
 scattering, calculated in \cite{ECL1,Weise1}, we get the width
 $\Gamma_{2p} = 2.675\,{\rm meV}$ of the excited $2p$ state and the
 yield $Y_{K^-d} = 1.90 \,\%$ of $X$--rays for the $K_{\alpha}$
 emission line of kaonic deuterium. The results, obtained in this
 paper, can be used for the planning of experiments on the
 measurements of the energy level displacement of the ground state of
 kaonic deuterium, caused by strong low--energy interactions. 
  \end{abstract}

\maketitle

\subsection{1. Introduction}

The consistent analysis of the complex S--wave scattering length
$\tilde{a}^{(0)}_{K^-d}$ of $K^-d$ scattering has been carried out in
\cite{Oset} by means of the solution of the Faddeev equations in the
fixed centre approximation.  The complex S--wave scattering length
$\tilde{a}^{(0)}_{K^-d}$ of $K^-d$ scattering has been expressed in
terms of the complex S--wave scattering lengths of $\bar{K}N$
scattering \cite{Oset} as follows
\begin{eqnarray}\label{label1}
\tilde{a}^{(0)}_{K^-d} = \frac{m_d}{m_K + m_d}\int
d^3x\,|\Phi_d(\vec{r}\,)|^2\,\hat{A}^{(0)}_{K^-d}(r),
\end{eqnarray}
where $\Phi_d(\vec{r}\,)$ is the wave function of the deuteron in the
ground state \cite{WFD}. The complex function
$\hat{A}^{(0)}_{K^-d}(r)$ is defined by $\hat{A}^{(0)}_{K^-d}(r) =
\hat{A}^{(0)}_p(r) + \hat{A}^{(0)}_n(r)$. The functions
$\hat{A}^{(0)}_p(r)$ and $\hat{A}^{(0)}_n(r)$ are the solutions of the
Faddeev equations in the fixed centre approximation. They are equal to
\cite{Oset}
\begin{eqnarray}\label{label2}
  \hspace{-0.3in}\hat{A}^{(0)}_p(r) &=&\frac{\displaystyle
  \hat{a}^{(0)}_p + \frac{\hat{a}^{(0)}_p\hat{a}^{0(0)}_n}{r} +
  \frac{\hat{a}^{(0)}_p\hat{a}^{(0)}_n -(\hat{a}^{(0)}_x)^2}{r} +
  \frac{\hat{a}^{(0)}_n(\hat{a}^{(0)}_p\hat{a}^{(0)}_n
  -(\hat{a}^{(0)}_x)^2)}{r^2}}{\displaystyle 1 +
  \frac{\hat{a}^{0(0)}_n}{r} -
  \frac{\hat{a}^{(0)}_p\hat{a}^{(0)}_n}{r^2} -
  \frac{\hat{a}^{(0)}_n(\hat{a}^{(0)}_p\hat{a}^{0(0)}_n
  -(\hat{a}^{(0)}_x)^2)}{r^3} },\nonumber\\
 \hspace{-0.3in}\hat{A}^{(0)}_n(r) &=& \hat{a}^{(0)}_n +
  \frac{\hat{a}^{(0)}_n}{r}\,\hat{A}^{(0)}_p(r) \quad,\quad
  \hat{A}^{x(0)}_n(r) = \frac{\displaystyle \hat{a}^{(0)}_x +
  \frac{\hat{a}^{(0)}_x\hat{a}^{(0)}_n}{r} +
  \frac{\hat{a}^{(0)}_x\hat{a}^{(0)}_n}{r^2}\,\hat{A}^{(0)}_p(r)}{\displaystyle
  1 + \frac{\hat{a}^{0(0)}_n}{r}},
\end{eqnarray}
where the complex S--wave scattering lengths of $\bar{K}N$ scattering
$\hat{a}^{(0)}_p$, $\hat{a}^{(0)}_n$, $\hat{a}^{(0)}_x$ and
$\hat{a}^{0{(0)}}_n $ are defined by \cite{Oset}
\begin{eqnarray}\label{label3}
\hat{a}^{(0)}_p &=& \Big(1 +
\frac{m_K}{m_N}\Big)\tilde{a}_{K^-p}(K^-p)\;,\;\hat{a}^{(0)}_n =
\Big(1 + \frac{m_K}{m_N}\Big)\tilde{a}_{K^-n}(K^-n),\nonumber\\
\hat{a}^{(0)}_x &=& \Big(1 +
\frac{m_K}{m_N}\Big)\tilde{a}_{K^-p}(\bar{K}^0n)\;,\;
\hat{a}^{0{(0)}}_n = \Big(1 +
\frac{m_K}{m_N}\Big)\tilde{a}_{\bar{K}^0n}(\bar{K}^0n).
\end{eqnarray}
According to \cite{Oset}, the Faddeev equations for the complex
S--wave scattering length of $K^-d$ scattering length take the form
\begin{eqnarray}\label{label4}
  T^{(0)}_p&=&t^{(0)}_p + t^{(0)}_p G_0 T^{(0)}_n + t^{x(0)}_p G_0
  T^{x(0)}_n\,, \nonumber\\ T^{(0)}_n&=&t^{(0)}_n + t^{(0)}_n G_0
  T^{(0)}_p\,, \nonumber\\ T^{x(0)}_n&=&t^{x(0)}_n + t^{0(0)}_n G_0
  T^{x(0)}_n + t^{x(0)}_n G_0 T^{(0)}_n.
\end{eqnarray}
As has been pointed out in \cite{Oset}, these equations describe pure
elastic and charge--exchange processes and require as input only the
amplitudes and propagators, where $G_0$ is the free kaon propagator
and $t^{(0)}_p$ and $t^{(0)}_n$ are the $\mathbb{T}$--matrices for
$K^-p$ and $K^-n$ elastic scattering, respectively, and $t^{0(0)}_n$
is the $\mathbb{T}$--matrix of $\bar{K}^0n$ scattering $\bar{K}^0 n
\to \bar{K}^0n$. For the proton partition $T^{(0)}_p$ there is also a
contribution from the charge-exchange channel $K^- p \to \bar{K}^0 n $
with the elementary $\mathbb{T}$-matrices $T^{x(0)}_p$ and
$T^{x(0)}_n$, describing $\bar{K}^0 n n \to K^- p n$ transition
including the multiple rescattering in the intermediate inelastic
states. In the approximation, proposed in \cite{Oset}, $t^{x(0)}_p =
t^{x(0)}_n$ and, correspondingly, $T^{x(0)}_n = T^{x(0)}_p$.

In this paper following the technique, developed in \cite{Oset}, we
derive the Faddeev equations for the P--wave amplitude of $K^-d$
scattering. We solve these equations in the fixed centre approximation
and calculate the complex P--wave scattering length of $K^-d$
scattering. We express the energy level displacement of the excited
$np$ state of kaonic deuterium in terms of the complex P--wave
scattering length of $K^-d$ scattering.

The paper is organised as follows. In section 2 we derive the Faddeev
equations for the P--wave amplitude of $K^-d$ scattering. Solving the
Faddeev equations in the fixed centre approximation we obtain the
complex P--wave scattering length of $K^-d$ scattering in terms of the
complex S--wave and P--wave scattering lengths of $\bar{K}N$
scattering. The numerical values of the complex S--wave and P--wave
scattering lengths of $\bar{K}N$ scattering are given in section 3.
The calculation of the complex S--wave and P--wave scattering lengths
of $\bar{K}N$ scattering is carried out within the $SU(3)$
coupled--channel approach and chiral Lagrangians with derivative
meson--baryon couplings invariant under chiral $SU(3)\times SU(3)$
symmetry.  In section 4 we calculate the numerical values of the
complex S--wave and P--wave scattering lengths of $K^-d$ scattering
and the energy level displacement of the kaonic deuterium in the
excited $np$ state, caused by strong low--energy interactions.  We
give the numerical values for the S--wave and P--wave scattering
lengths of $K^-d$ scattering and the energy level displacements of the
ground $1s$ state and the excited $2p$ state of kaonic deuterium. The
complex S--wave scattering length of $K^-d$ scattering is in
reasonable agreement with the results, obtained in \cite{Oset}.  In
section 5 using our prediction for the width $\Gamma_{2p} =
10.203\,{\rm meV}$ of the excited $2p$ state of kaonic deuterium and
the quantum--classical Monte Carlo cascade model, developed in
\cite{Faifman}, we calculate the yield $Y_{K^-d} = 0.27\,\%$ of
$X$--rays for the $K_{\alpha}$ emission line of kaonic deuterium. In
Conclusion we discuss the obtained results and the estimate of an
uncertainty of our solution for the complex P--wave scattering length
of $K^-d$ scattering, which is of about $15\,\%$. We calculate the
complex S--wave and P--wave scattering lengths of $K^-d$ scattering
and the energy level displacements of the ground $1s$ and excited $2p$
state of kaonic deuterium for the complex S--wave and P--wave
scattering lengths, obtained in \cite{ECL1,Weise1}. We get the
following values for the width $\Gamma_{2p} = 2.675\,{\rm meV}$ of the
excited $2p$ state and the yield $Y_{K^-d} = 1.90\,\%$ of $X$--rays
for the $K_{\alpha}$ emission line of kaonic deuterium. In Appendix A
we give a detailed calculation of the contributions of the single
({\it impulse}) and double scattering to the complex P--wave
scattering length of $K^-d$ scattering. In Appendix B we outline the
calculation of the complex S--wave and P--wave scattering lengths of
$\bar{K}N$ scattering.

\subsection{2. Faddeev equations for P--wave amplitude  of $K^-d$ scattering and 
 P--wave scattering length of $K^-d$ scattering in the fixed centre
 approximation}

For the P--wave scattering $\mathbb{T}$--matrices we use the index
$(1)$. This defines $t^{(1)}_p$, $t^{(1)}_n$, $t^{x(1)}_n$,
$t^{0(1)}_n$, $T^{(1)}_p$, $T^{(1)}_n$ and $T^{x(1)}_n$,
respectively. In this notation the Faddeev equations for the P--wave
amplitude of $K^-d$ scattering read
\begin{eqnarray}\label{label5}
  T^{(1)}_p&=&t^{(1)}_p + t^{(1)}_p G_0 T^{(0)}_n + t^{(0)}_p G_0
  T^{(1)}_n + t^{x(1)}_p G_0 T^{x(0)}_n + t^{x(0)}_p G_0 T^{x(1)}_n\,,
  \nonumber\\ T^{(1)}_n&=&t^{(1)}_n + t^{(1)}_n G_0 T^{(0)}_p +
  t^{(0)}_n G_0 T^{(1)}_p\,, \nonumber\\ T^{x(1)}_n&=&t^{x(1)}_n +
  t^{0(1)}_n G_0 T^{x(0)}_n + t^{0(0)}_n G_0 T^{x(1)}_n + t^{x(1)}_n
  G_0 T^{(0)}_n + t^{x(0)}_n G_0 T^{(1)}_n.
\end{eqnarray}
In the fixed centre approximation the Faddeev equations
Eq.(\ref{label5}) reduce to the system of algebraical equations for
the amplitudes $T^{(1)}_p \to A^{(1)}_p(r)$, $T^{(1)}_n \to
A^{(1)}_n(r)$ and $T^{x(1)}_n \to A^{x(1)}_n(r)$
\begin{eqnarray}\label{label6}
  \hspace{-0.3in}\hat{A}^{(1)}_p(r)&=&\hat{a}^{(1)}_p + \frac{1}{6}\,\hat{a}^{(1)}_p
  \frac{1}{r} \hat{A}^{(0)}_n(r) + \frac{1}{6}\,\hat{a}^{(0)}_p
  \frac{1}{r} \hat{A}^{(1)}_n(r) - \frac{1}{6}\,\hat{a}^{(1)}_x
  \frac{1}{r} \hat{A}^{x(0)}_n(r) - \frac{1}{6}\,\hat{a}^{(0)}_x
  \frac{1}{r} \hat{A}^{x(1)}_n(r)\,,
  \nonumber\\ 
\hspace{-0.3in}\hat{A}^{(1)}_n(r)&=&\hat{a}^{(1)}_n +
  \frac{1}{6}\,\hat{a}^{(1)}_n \frac{1}{r} \hat{A}^{(0)}_p(r) +
  \frac{1}{6}\,\hat{a}^{(0)}_n \frac{1}{r} \hat{A}^{(1)}_p(r)\,,
  \nonumber\\ 
\hspace{-0.3in}\hat{A}^{x(1)}_n(r)&=&\hat{a}^{(1)}_x -
  \frac{1}{6}\,\hat{a}^{0(1)}_n \frac{1}{r} \hat{A}^{x(0)}_n(r) -
  \frac{1}{6}\,\hat{a}^{0(0)}_n \frac{1}{r} \hat{A}^{x(1)}_n(r) +
  \frac{1}{6}\,\hat{a}^{(1)}_x \frac{1}{r} \hat{A}^{(0)}_n(r) +
  \frac{1}{6}\,\hat{a}^{(0)}_x \frac{1}{r} \hat{A}^{(1)}_n(r),
\end{eqnarray}
where $\hat{a}^{(1)}_p$, $\hat{a}^{(1)}_n$ and $\hat{a}^{(1)}_x$ are
the complex P--wave scattering length of $\bar{K}N$ scattering,
defined by analogy with $a^{(0)}_p$, $a^{(0)}_n$ and $a^{(0)}_x$ of
Eq.(\ref{label3}).  The P--wave scattering length of $K^-d$ scattering
is equal
\begin{eqnarray}\label{label7}
\tilde{a}^{(1)}_{K^-d} = \frac{m_d}{m_K + m_d}\int
d^3x\,|\Phi_d(\vec{r}\,)|^2\,\hat{A}^{(1)}_{K^-d}(r),
\end{eqnarray}
where $\hat{A}^{(1)}_{K^-d}(r) = \hat{A}^{(1)}_p(r) +
\hat{A}^{(1)}_n(r)$.  The amplitudes $\hat{A}^{(1)}_p(r)$ and
$\hat{A}^{(1)}_n(r)$ are the solutions of Eq.(\ref{label6}). They are
equal to
\begin{eqnarray}\label{label8}
  \hspace{-0.3in}&&\hat{A}^{(1)}_p(r)\Big(1 +
  \frac{1}{6}\frac{\hat{a}^{0(0)}_n}{r} -
  \frac{1}{36}\frac{\hat{a}^{(0)}_n\hat{a}^{(0)}_p}{r^2} -
  \frac{1}{216}\frac{\hat{a}^{(0)}_n(\hat{a}^{(0)}_p\hat{a}^{0(0)}_n -
  (\hat{a}^{(0)}_x)^2)}{r^3}\Big) = \hat{a}^{(1)}_p +
  \frac{1}{6}\frac{\hat{a}^{(1)}_p\hat{a}^{0(0)}_n}{r} +
  \frac{1}{6}\frac{\hat{a}^{(1)}_n\hat{a}^{(0)}_p}{r} -
  \frac{1}{6}\frac{\hat{a}^{(1)}_x\hat{a}^{(0)}_x}{r}\nonumber\\
\hspace{-0.3in}&& +
  \frac{1}{36}\,\frac{\hat{a}^{(1)}_n(\hat{a}^{(0)}_p \hat{a}^{0(0)}_n
  - (\hat{a}^{(0)}_x)^2)}{r^2} +
  \frac{1}{6}\frac{\hat{a}^{(1)}_p}{r}\hat{A}^{(0)}_n(r) -
  \frac{1}{6}\frac{\hat{a}^{(1)}_x}{r}\hat{A}^{x(0)}_n(r) +
  \frac{1}{36}\frac{\hat{a}^{(1)}_p\hat{a}^{0(0)}_n
  }{r^2}\hat{A}^{(0)}_n(r) +
  \frac{1}{36}\frac{\hat{a}^{(1)}_n\hat{a}^{(0)}_p
  }{r^2}\hat{A}^{(0)}_p(r)\nonumber\\
\hspace{-0.3in}&& + \frac{1}{36}\frac{\hat{a}^{0(1)}_n\hat{a}^{(0)}_x
}{r^2}\hat{A}^{x(0)}_n(r) -
\frac{1}{36}\frac{\hat{a}^{(1)}_x\hat{a}^{0(0)}_n
}{r^2}\hat{A}^{x(0)}_n(r) -
\frac{1}{36}\frac{\hat{a}^{(1)}_x\hat{a}^{(0)}_x
}{r^2}\hat{A}^{(0)}_n(r) +
\frac{1}{216}\frac{\hat{a}^{(1)}_n(\hat{a}^{(0)}_p \hat{a}^{0(0)}_n -
  (\hat{a}^{(0)}_x)^2)}{r^3}\hat{A}^{(0)}_p(r),\nonumber\\
\hspace{-0.3in}&&\hat{A}^{(1)}_n(r) = \hat {a}^{(1)}_n +
\frac{1}{6}\frac{\hat{a}^{(1)}_n}{r}\hat{A}^{(0)}_p(r) +
\frac{1}{6}\frac{\hat{a}^{(0)}_n}{r}\hat{A}^{(1)}_p(r),
\end{eqnarray}
where $\hat{A}^{(0)}_p(r)$, $\hat{A}^{(0)}_n(r)$ and
$\hat{A}^{x(0)}_n(r)$ are solutions of the Faddeev equations in the
fixed centre approximation for the complex S--wave scattering length
of $K^-d$ scattering Eq.(\ref{label2}) \cite{Oset}.

Expanding the amplitudes $\hat{A}^{(1)}_p(r)$ and $\hat{A}^{(1)}_n(r)$
in powers of $1/r$ and keeping only the terms of order of $1/r$ one
arrives at the complex P--wave scattering length of $K^-d$ scattering
in the single and double scattering approximation
\begin{eqnarray}\label{label9}
\hspace{-0.3in}\tilde{a}^{(1)}_{K^-d} = \frac{m_d}{m_K +
m_d}\Big(\hat{a}^{(1)}_p + \hat{a}^{(1)}_n +
\frac{1}{3}(\hat{a}^{(1)}_p \hat{a}^{(0)}_n + \hat{a}^{(1)}_n
\hat{a}^{(0)}_p - \hat{a}^{(1)}_x \hat{a}^{(0)}_x)\int
\frac{d^3x}{r}\,|\Phi_d(\vec{r}\,)|^2 + \ldots\Big).
\end{eqnarray} 
This result is confirmed in Appendix A by a direct calculation in the
effective low--energy quantum field theory.

\subsection{3. Complex S--wave and P--wave scattering lengths of 
$\bar{K}N$ scattering}

For the evaluation of the numerical value of the complex P--wave
scattering length of $K^-d$ scattering and the energy level
displacement of the excited $np$ state of kaonic deuterium, we have
to calculate the numerical values of the complex S--wave and P--wave
scattering lengths of $\bar{K}N$ scattering. A detailed procedure of
the calculation of the complex S--wave and P--wave scattering lengths
is expounded in Appendix B. The numerical values of them are equal to
\begin{eqnarray}\label{label10}
 \tilde{a}^{(0)}_{K^-p}(K^-p) &=& -\,0.680 + i\,0.639\,{\rm
 fm}\quad,\quad \tilde{a}^{(1)}_{K^-p}(K^-p) = -\,0.069 +
 i\,0.179\,{\rm fm^3},\nonumber\\ \tilde{a}^{(0)}_{K^-p}(\bar{K}^0n)
 &=& +\,0.980 - i\,0.543\,{\rm fm}\quad,\quad
 \tilde{a}^{(1)}_{K^-p}(\bar{K}^0n) = -\,0.053 + i\,0.176\,{\rm
 fm^3},\nonumber\\ \tilde{a}^{(0)}_{K^-n}(K^-n) &=& +\,0.300 +
 i\,0.096\,{\rm fm}\quad,\quad \tilde{a}^{(1)}_{K^-n}(K^-n) = -\,0.122
 + i\,0.355\,{\rm fm^3},\nonumber\\
 \tilde{a}^{(0)}_{\bar{K}^0n}(\bar{K}^0n) &=& -\,0.680 +
 i\,0.639\,{\rm fm}\quad,\quad
 \tilde{a}^{(1)}_{\bar{K}^0n}(\bar{K}^0n) = -\,0.069 + i\,0.179\,{\rm
 fm^3}.
\end{eqnarray}
We have calculated the complex S--wave and P--wave scattering lengths
of $\bar{K}N$ scattering within the $SU(3)$ coupled--channel approach
\cite{ECL1}, chiral dynamics with chiral $SU(3)\times SU(3)$ invariant
low--energy meson--baryon interactions with derivative couplings
\cite{ECL1}--\cite{ECL3} and the account for the contributions of
baryon resonances \cite{IV5} and scalar meson resonances
\cite{Jaffe,Ecker}. The chiral Lagrangian of low--energy
interactions of the ground--state baryon octet $B(x)$ with octet of
pseudoscalar mesons $P(x)$ invariant under $SU(3)\times
SU(3)$ chiral symmetry is \cite{ECL1}
\begin{eqnarray}\label{label11}
 \hspace{-0.3in}&&{\cal L}(x) =
 \langle\bar{B}(x)(i\gamma^{\mu}\partial_{\mu} - m_0)B(x)\rangle +
 \langle \bar{B}(x)i\gamma^{\mu}[s_{\mu}(x),B(x)]\rangle -\,g_A\,(1 -
 \alpha_D) \langle \bar{B}(x)\gamma^{\mu}[p_{\mu}(x),B(x)]\rangle\nonumber\\
 \hspace{-0.3in} && + \alpha_D\, \langle
 \bar{B}(x)\gamma^{\mu}\{p_{\mu}(x),B(x)\}\rangle + \frac{1}{4}\, b_D
 \langle \bar{B}(x)\{\chi_+(x),B(x)\}\rangle + \frac{1}{4}\, b_F
 \langle \bar{B}(x)[\chi_+(x),B(x)]\rangle + \frac{1}{4}\, b_0 \langle
 \bar{B}(x)\langle \chi_+(x)\rangle B(x)\rangle\nonumber\\
 \hspace{-0.3in} &&+ \frac{1}{2}\, d_1 \langle
  \bar{B}(x)\{p_{\mu}(x),[p^{\mu}(x),B(x)]\}\rangle + \frac{1}{2}\, d_2
  \langle \bar{B}(x) [p_{\mu}(x),[p^{\mu}(x),B(x)]]\rangle+ \frac{1}{2}\, d_3 \langle
  \bar{B}(x)p_{\mu}(x)\rangle \langle p^{\mu}(x) B(x)\rangle\nonumber\\
  \hspace{-0.3in}&& + \frac{1}{2}\, d_4 \langle \bar{B}(x)\langle
  p_{\mu}(x) p^{\mu}(x)\rangle B(x)\rangle + \ldots\nonumber\\
  \hspace{-0.3in}&&s_{\mu}(x) =
  \frac{1}{2}\,[U^{\dagger}(x),\partial_{\mu}U(x)]\;,\; p_{\mu}(x) =
  \frac{1}{2i}\,\{U^{\dagger}(x)\partial_{\mu}U(x)\}\;,\; \chi_+(x) =
  2B_0(U^{\dagger}(x){\cal M}U^{\dagger}(x) + U(x){\cal M}U(x)),
\end{eqnarray}
where $B(x) = (N, \Lambda^0, \Sigma, \Xi)$ is the ground--state baryon
octet \cite{PDG10}, $U^2(x) = e^{\textstyle\, \sqrt{2}\,i \gamma^5
P(x)/F_{\pi}}$, $P(x) = (\pi, \eta, K, \bar{K})$ and $F_{\pi} =
92.4\,{\rm MeV}$ are the octet of low--lying pseudoscalar mesons and
the PCAC constant \cite{Adler}, $\langle \ldots\rangle$ are the traces
over the $SU(3)$ indices, $g_A = 1.275$ \cite{Abele1,Faber3},
$\alpha_D = 0.635$ and $m_0$ is the baryon mass for current quark
masses zero; ${\cal M} = {\rm diag}(m_u,m_d,m_s)$ is a diagonal
$3\times 3$ matrix with current quark masses $m_q$ for $q = u, d$ and
$s$, respectively, and $B_0 = - \langle \bar{q}q\rangle/F^2_{\pi}$ ,
where $\langle \bar{q}q\rangle$ is the quark condensate. The current
quark masses and the quark condensate are defined at the normalisation
scale $\mu = 1\,{\rm GeV}$ \cite{Gasser}. The ellipsis denotes the
contributions of the derivative couplings of the $\Lambda(1405)$
resonance, the baryon decuplet $\underline{\bf 10} = (\Delta,
\Sigma^*,\ldots)$ and other baryon resonances \cite{IV5} with quantum
numbers $J^P = \frac{1}{2}^+$ , belonging to octets of $SU(3)_f$
symmetry \cite{PDG10}, with octet of pseudoscalar mesons and the
ground--state baryon octet invariant under chiral $SU(3)\times SU(3)$
symmetry, and also chiral $SU(3)\times SU(3)$ invariant interactions
of the ground--state baryon octet with the nonet of scalar meson
resonances \cite{Jaffe,Ecker} (see Appendix B). The parameters $b_D$
and $b_F$ define the mass--splitting of the ground--state baryons
$m_{\Sigma} - m_{\Lambda^0} = \frac{4}{3}\,b_D(m^2_K - m^2_{\pi})$ and
$m_{\Xi} - m_{\Sigma} = (b_D + b_F)\,(m^2_{\pi} - m^2_K)$. They are
equal to $b_D = + 0.051\,{\rm fm}$ and $b_F = -0.158\,{\rm fm}$. The
parameter $b_0$ determines the $\sigma_{\pi N}$--term of $\pi N$
scattering $2\sigma_{\pi N} = - m^2_{\pi} (2 b_0 + b_D + b_F)$. It is
equal to $b_0 = - 0.561\,{\rm fm}$, calculated in terms of the
experimental value $\sigma^{(\exp)}_{\pi N} = 61\,{\rm MeV}$
\cite{PSI}, which agrees well with the theoretical one $\sigma^{(\rm
th)}_{\pi N} = 60\,{\rm MeV}$ \cite{IV99}. The amplitudes $M$ of
low--energy $\bar{K}N$ scattering are determined in the $SU(3)$
coupled--channel approach by the matrix equation $M^{-1} = M^{-1}_0 -
G$, where $M_0$ are the amplitudes of $\bar{K}N$ scattering,
calculated with the chiral Lagrangian Eq.(\ref{label14}) and other
Lagrangians, adduced in Appendix B, in the tree--approximation
\cite{ECL1}.  Since we are interested in the scattering lengths we
calculate the matrix elements of the diagonal matrix $G$, given by the
meson--baryon loop diagrams, in the non--relativistic approximation
within the dimensional regularisation. As a result the matrix elements
are imaginary and proportional to $k$ and $k^3$ for the S--wave and
P--wave $\bar{K}N$ scattering, respectively, for kinematically opened
channels, where $k$ is a momentum transfer.  The complex S--wave
scattering length of $K^-p$ scattering we set equal to the preliminary
experimental value by the SIDDHARTA Collaboration \cite{SMI}. In our
approach the imaginary parts of the complex S--wave and P--wave
scattering lengths of $\bar{K}N$ scattering are defined by the
contributions of the $\Lambda(1405)$ and $\Sigma(1385)$
resonances. This agrees well with the analysis of low--energy
$\bar{K}N$ interactions in the S--wave and P--wave states, carried out
for the investigation of the properties of antikaon--nuclear
quasibound states in \cite{Weise1}. For the coupling constants
$d_j\,(j = 1,2,3,4)$, which are input parameters, we have got the
following values: $d_1 = -\,0.389\,{\rm fm}$, $d_2 = -\,0.709\,{\rm
fm}$, $d_3 = +\,2.816\,{\rm fm}$ and $d_4 = -\,0.619\,{\rm fm}$. As
has been found the contribution of the scalar meson resonances is not
essential for reasonable values of coupling constant of the
interactions of scalar meson resonances with the ground--state baryons
(see Appendix B).

\subsection{4. Energy level shift and width of excited $np$ state of
 kaonic deuterium}

Following \cite{IV5} we define the shift and width of the energy level
of the excited $np$ state of kaonic deuterium, where $n$ is the {\it
principal} quantum number, in terms of the complex P--wave scattering
length $\tilde{a}^{(1)}_{K^-d}(K^-d)$ of $K^-d$ scattering. We get
\begin{eqnarray}\label{label12}
  \epsilon_{np} &=& - 2\,\frac{\alpha^5}{n^3}\, \Big(1 -
\frac{1}{n^2}\Big)\, \Big(\frac{m_Km_d}{m_K + m_d}\Big)^4\,{\rm
Re}\,\tilde{a}^{(1)}_{K^-d},\nonumber\\ \Gamma_{np} &=&
4\,\frac{\alpha^5}{n^3}\, \Big(1 - \frac{1}{n^2}\Big)\,
\Big(\frac{m_Km_d}{m_K + m_d}\Big)^4\,{\rm
Im}\,\tilde{a}^{(1)}_{K^-d},
\end{eqnarray}
where $\alpha = 1/137.036$ is the fine--structure constant. 

Using the numerical values of the complex S--wave and P--wave
scattering lengths, evaluated in section 3, we obtain the following
numerical values of the S--wave and P--wave scattering lengths of
$K^-d$ scattering
\begin{eqnarray}\label{label13}
 \tilde{a}^{(0)}_{K^-d} &=& -\,1.273 + i\,2.435\,{\rm fm},\nonumber\\
 \tilde{a}^{(1)}_{K^-d} &=& -\,0.352 + i\,0.432\,{\rm fm^3}.
\end{eqnarray}
They give the following energy level displacements of the ground $1s$
and excited $2p$ states of kaonic deuterium
\begin{eqnarray}\label{label14}
 \epsilon_{1s} &=& 0.766\,{\rm keV}\quad,\quad \Gamma_{1s} =
 2.933\,{\rm keV},\nonumber\\ \epsilon_{2p} &=& 4.158\,{\rm
 meV}\quad,\quad \Gamma_{2p} = 10.203\,{\rm meV}.
\end{eqnarray}
The numerical value of the complex S--wave scattering length of $K^-d$
scattering agrees reasonably well with the results, obtained in
\cite{Oset}. An uncertainty of the complex P--wave scattering length
of $K^-d$ scattering, which is estimated as $15\,\%$, we discuss in
the Conclusion.

\subsection{5. Yield of $X$--rays for $K_{\alpha}$ emission line of  kaonic deuterium}

The results of the calculation of the yields of $X$-rays of the
$K_{\alpha}$ emission lines for kaonic hydrogen and deuterium depend
considerably on the values of the widths of the excited $2p$ state of
kaonic atoms \cite{Ericson}.  Using the calculation scheme based on
the quantum--classical Monte Carlo cascade model, developed in
\cite{Faifman}, we obtain the following yields of the $K_{\alpha}$
emission lines
\begin{eqnarray}\label{label15}
 Y_{K^-p}  &=& 1.80\,{\%}\,,\quad\quad \Gamma_{1p} = 1.979\, {\rm meV},\nonumber\\
 Y_{K^-d} &=& 0.27\,{\%}\,,\quad\quad \Gamma_{2p} = 10.203\,{\rm meV}
\end{eqnarray}
for kaonic hydrogen and deuterium, respectively.

Our result $\Gamma_{2p} = 1.979\,{\rm meV}$ for the width of the
excited $2p$ state of kaonic hydrogen agrees well with $\Gamma_{2p} =
2\,{\rm meV}$, obtained in \cite{IV5}. The theoretical value $Y_{K^-p}
= 1.80\,\%$ is in good agreement with the experimental one $Y_{K^-p} =
1.5(5)\,\%$ \cite{KEK}. The theoretical value $Y_{K^-d} = 0.27\,\%$ can
be used for the planning experiments on the measurement of the energy
level displacement of the ground $1s$ state of kaonic deuterium.

\subsection{6. Conclusion}

We have investigated the properties of exotic atom -- kaonic deuterium
in the excited $np$ state, where $n$ is the {\it principal} quantum
number, relative to strong low--energy interactions, described by
chiral Lagrangians with derivative meson--baryon couplings invariant
under chiral $SU(3)\times SU(3)$ symmetry. We have calculated the
energy level shift and width of the excited $np$ state in terms of the
complex P--wave scattering length of $K^-d$ scattering. Since $K^-d$
scattering is a three--body into three--body reaction, the most
appropriate tool for the investigation of the P--wave amplitude of
$K^-d$ scattering is the Faddeev equations \cite{Oset}.  Following
\cite{Oset}, where the Faddeev equations for the S--wave amplitude of
$K^-d$ scattering has been solved in the fixed centre approximation,
we have calculated the Faddeev equations for the P--wave amplitude of
low--energy $K^-d$ scattering in the fixed centre approximation. In
such an approximation the complex S--wave and P--wave scattering
lengths of $K^-d$ scattering are expressed in terms of the complex
S--wave and P--wave scattering lengths of $\bar{K}N$ scattering. The
calculation of the complex S--wave and P--wave scattering lengths of
$\bar{K}N$ scattering is carried out within the $SU(3)$
coupled--channel approach and chiral Lagrangians with derivative
meson--baryon couplings invariant under chiral $SU(3)\times SU(3)$
symmetry. The complex S--wave scattering length of $K^-p$ scattering
we have set equal to recent experimental value, measured by the
SIDDHARTA Collaboration.

We note that our result for the real part of the complex P--wave
scattering length of $K^-p$ scattering, obtained in this paper,
differs with a sign from that, calculated in \cite{IV5}. Such a
discrepancy is caused by different dynamics, which are used in
\cite{IV5} and in the present paper. Indeed, for the calculation of
the complex P--wave scattering length in the present paper we use
chiral Lagrangians with derivative meson--baryon couplings, derived
within non--linear realisation of chiral $SU(3)\times SU(3)$ symmetry
\cite{ECL2}, which contain also additional interactions with the
coupling constants $d_j$ for $(j =1,2,3,4)$ and $b_{\ell}$ for $\ell =
0,D,F$ \cite{ECL1}. In \cite{IV5} the calculation of the complex
P--wave scattering length of $K^-p$ scattering has been carried out
with chiral Lagrangians, derived within linear realisation of chiral
$SU(3)\times SU(3)$ symmetry. These chiral Lagrangians do not contain
also the interactions with the coupling constants $d_j$ for $(j
=1,2,3,4)$ and $b_{\ell}$ for $\ell = 0,D,F$, which are specific for
non--linear realisation of chiral symmetry and have no analogy within
its linear realisation. Since the imaginary part of the complex
P--wave scattering length is defined by the dominant contribution of
the $\Sigma(1385)$ resonance, the values of the imaginary parts,
calculated in the present paper and in \cite{IV5}, agree well.

The numerical value of the complex S--wave scattering length of $K^-d$
scattering Eq.(\ref{label13}) agrees reasonably well with the results
obtained in \cite{Oset}.  We note that the complex S--wave scattering
length of $K^-d$ scattering has been also investigated within the
effective field theory approach \cite{KDS1,KDS2}. In \cite{KDS1} the
solution of the Faddeev equations, obtained in the fixed centre
approximation, for the complex S--wave scattering length has been
confirmed within the effective field theory approach. In \cite{KDS2}
the effective field theory approach has been applied to the
calculation of the nucleon recoil corrections to the double scattering
contribution to the complex S--wave scattering length of $K^-d$
scattering, obtained in the fixed centre approximation.  As has been
found in \cite{KDS2}, the nucleon recoil corrections make up of about
$(10-15)\,\%$.

Now let us discuss an uncertainty of the proposed solution of the
Faddeev equations for the complex P--wave scattering length of $K^-d$
scattering, obtained in the fixed centre approximation. As has been
pointed out by Gal \cite{Gal06}, an uncertainty of the solution of the
Faddeev equations for the complex S--wave scattering length of $K^-d$
scattering, calculated in the fixed centre approximation, is of about
$(10-25)\,\%$. Such an estimate has been deduced from the comparison
of the solution, found in \cite{Oset}, with other solutions of the
Faddeev equations, applied to the calculation of the complex S--wave
scattering length of $K^-d$ scattering \cite{KDS3}--\cite{KDS5}. Of
course, the lack of the experimental data on the complex S--wave
scattering length of $K^-d$ scattering does not allow to understand a
real uncertainty of theoretical schemes.

Since the calculation of the complex P--wave scattering length of
$K^-d$ scattering has not been yet carried out in literature, and our
paper is the first attempt of this kind, we have no possibility to
compare our solution with any others. Thus, for the estimate of the
theoretical uncertainty of our solution of the Faddeev equations for
the complex P--wave scattering length of $K^-d$ scattering we follow
the results, obtained in \cite{KDS2}. As has been shown in \cite{KDS2},
the nucleon recoil correction to the double scattering contribution
of the static solution of the Faddeev equations makes up of about
$15\,\%$. Since we neglect the nucleon recoil, one can accept $15\,\%$
as an uncertainty of our static solution of the Faddeev equations.

Such an estimate of an uncertainty, applied to the total solution of
the Faddeev equations in the fixed centre approximation, can be
supported by a convergence of the expansion of our solution for the
complex P--wave scattering length of $K^-d$ scattering in powers of
$1/r$. As we have shown in Appendix A, the contribution of the single
and double scattering $\tilde{a}^{(1)}_{K^-d} =
(\tilde{a}^{(1)}_{K^-d})_{\rm s.sc.} + (\tilde{a}^{(1)}_{K^-d})_{\rm
  d.sc.}  = - 0.262\, + i\, 0.548\,{\rm fm^3}$ dominates in the
complex P--wave scattering length of $K^-d$ scattering.  The account
for the contribution of the triple scattering $\tilde{a}^{(1)}_{K^-d}
= (\tilde{a}^{(1)}_{K^-d})_{\rm s.sc.} + (\tilde{a}^{(1)}_{K^-d})_{\rm
  d.sc.} + (\tilde{a}^{(1)}_{K^-d})_{\rm tr.sc.} = - 0.277\, + i\,
0.525\,{\rm fm^3}$, obtained from the expansion of the exact solution
of the Faddeev equations, corroborates only such a dominance. The
contributions of higher multiple scattering are equal to $\delta
\tilde{a}^{(1)}_{K^-d} = - 0.075 - \,i\,0.072\,{\rm fm^3}$. They make
up of about $21\,\%$ and $17\,\%$ of the real and imaginary parts of
the total complex P--wave scattering length, respectively.  Of course,
a proof of the convergence by means of the calculation of the
contributions of higher $n$--multiple scattering for $n\ge 4$,
proportional to the higher powers of $1/r^{(n -1)}$, averaged with the
deuteron wave function $|\Phi_d(\vec{r}\,)|^2$ as $\int d^3x\,
|\Phi_d(\vec{r}\,)|^2/r^{(n -1)} = \langle 1/r^{(n -1)}\rangle$,
stumbles against the problem of the regularisation and renormalisation
of these averaged values \cite{PDS}. The solution of this problem goes
beyond the scope of this paper.  We are planning to carry out this
analysis in our forthcoming publication. We note that without
truncation the evaluation of the complex P--wave scattering length
Eq.(\ref{label8}), caused by the multiple--scattering, does not suffer
from divergences at $r \to 0$. This agrees well with the results,
obtained in \cite{Oset}.

Using the numerical value of the complex P--wave scattering length of
$K^-d$ scattering Eq.(\ref{label13}) we have calculated the width of
the excited $2p$ state $\Gamma_{2p} = 10.203\,{\rm meV}$. This result
plays an important role for the theoretical analysis of the yield
$Y_{K^-d}$ of $X$--rays of the $K_{\alpha}$ emission line of kaonic
deuterium. Using the quantum--classical Monte Carlo cascade model
\cite{Faifman} we have obtained $Y_{K^-d} = 0.27\,\%$ for the width
$\Gamma_{2p} = 10.203\,{\rm meV}$ of the excited $2p$ state of kaonic
deuterium, For the yield of $X$--rays of the $K_{\alpha}$ emission
line of kaonic hydrogen we have got the value $Y_{K^-p} = 1.80\,\%$,
which agrees well with the experimental data $Y_{K^-p} = 1.5(5)\,\%$
\cite{KEK}.

Concluding this discussion we would like to note that the complex
S--wave scattering length of elastic $K^-n$ scattering
$\tilde{a}^{(0)}_{K^-n}(K^-n)= 0.300 + i\,0.096\,{\rm fm}$ or the
complex S--wave scattering length of $\bar{K}N$ scattering in the state
with isospin $I = 1$, i.e. $\tilde{a}_{I = 1} =
\tilde{a}^{(0)}_{K^-n}(K^-n)$, calculated in our approach to
$\bar{K}N$ scattering (see Appendix B) and given in
Eq.(\ref{label12}), possesses a small imaginary part ${\rm
  Im}\,\tilde{a}^{(0)}_{K^-n}(K^-n) = 0.096\,{\rm fm}$. This does not
contradict some estimates of the complex S--wave scattering length of
$\bar{K}N$ scattering in the state with isospin $I = 1$, obtained in
\cite{KDS1} from the complex S--wave scattering length of $K^-d$
scattering.  Nevertheless, the theoretical analysis of $\bar{K}N$
scattering, carried out in \cite{ECL1,CSL1}, shows that the imaginary
part of the complex S--wave scattering length of elastic $K^-n$
scattering is commensurable with the imaginary part of the complex
S--wave scattering length of elastic $K^-p$ scattering. As has been
found in \cite{ECL1}, the complex S--wave scattering length of elastic
$K^-n$ scattering is equal to $\tilde{a}^{(0)}_{K^-n}(K^-n)= 0.49 +
i\,0.70\,{\rm fm}$, which agrees well with the empirical result
$\tilde{a}^{(0)}_{K^-n}(K^-n)= 0.37+ i\,0.60\,{\rm fm}$, obtained in
\cite{Martin}, and the theoretical estimates in \cite{CSL1}. Thus, for
the completeness of our numerical predictions for the complex P--wave
scattering length of $K^-d$ scattering we take into account the
complex S--wave scattering lengths of $\bar{K}N$ scattering,
calculated in \cite{ECL1} and the complex P--wave scattering lengths
of $\bar{K}N$ scattering, calculated in \cite{Weise1}. We get
\begin{eqnarray}\label{label16}
 \tilde{a}^{(0)}_{K^-d} &=& -\,1.951 + i\,0.996\,{\rm fm},\nonumber\\
 \tilde{a}^{(1)}_{K^-d} &=& -\,0.174 + i\,0.113 \,{\rm fm^3}.
\end{eqnarray}
The calculation is performed for $a^{(0)}_{I = 0}(\bar{K}N) = -
\,1.63+ i\,0.42\,{\rm fm}$ and $a^{(0)}_{I = 1}(\bar{K}N) = 0.49 +
i\,0.70\,{\rm fm}$ \cite{ECL1} and $a^{(1)}_{I = 0}(\bar{K}N) = 0$ and
$a^{(1)}_{I = 1}(\bar{K}N) = - 0.114 + i\,0.098\,{\rm fm^3}$
\cite{Weise1}. The energy level displacements of the ground $1s$ and
excited $2p$ states of kaonic deuterium, calculated in terms of the
complex S--wave and P--wave scattering lengths Eq.(\ref{label16}), are
equal to
\begin{eqnarray}\label{label17}
 \epsilon_{1s} &=& 1.175\,{\rm keV}\quad,\quad \Gamma_{1s} =
 1.200\,{\rm keV},\nonumber\\ \epsilon_{2p} &=& 2.053\,{\rm
   meV}\quad,\quad \Gamma_{2p} = 2.675\,{\rm meV}.
\end{eqnarray}
In this case the yield of the $X$--rays of the $K_{\alpha}$ emission
line for kaonic deuterium is 
\begin{eqnarray}\label{label18}
 Y_{K^-d} &=& 1.90\,{\%}\,,\quad\quad \Gamma_{2p} = 2.675\,{\rm meV}.
\end{eqnarray}
Our results for the yields of $X$--rays of the $K_{\alpha}$ emission
line and the energy level displacements of kaonic deuterium in the
ground $1s$ state, calculated for the complex S--wave and P--wave
scattering lengths of $\bar{K}N$ scattering, obtained in this paper
and in \cite{ECL1,CSL1,Weise1}, can be used for the planning of
experiments on the measurements of the energy level displacement of
the ground $1s$ state of kaonic deuterium. The results, obtained in
this paper, can be also used by the SIDDHARTA Collaboration, measuring
currently the energy level displacement of the ground $1s$ state of
kaonic deuterium \cite{SIDDHARTA1,SIDDHARTA2}.

\subsection{7. Acknowledgement}

We are grateful to T. Ericson for numerous fruitful discussions and
comments during the work on the problems, expounded in this paper. We
acknowledge encouraging discussions with S. Kamalov.  The work of
A. I. and M. P. was supported by the Austrian ``Fonds zur F\"orderung
der Wissenschaftlichen Forschung'' (FWF) under contract P19487-N16 and
in part by the U.S. Department of Energy contract
No. DE-FG02-08ER41531, No. DE-AC02-06CH11357 and the Wisconsin Alumni
Research Foundation.

\subsection{Appendix A: Complex P--wave scattering length of $K^-d$ scattering. 
Single ({\it impulse}) and double scattering contributions}
\renewcommand{\theequation}{A-\arabic{equation}}
\setcounter{equation}{0}

In this Appendix we give a detailed calculation of the complex P--wave
scattering length of $K^-d$ scattering, keeping the contributions of
the single ({\it impulse}) and double scattering only. The P--wave
amplitude $M^{(1)}(K^-d \to K^-d)$ of low--energy elastic $K^-d$
scattering relates to the complex P--wave scattering length
$a^{(1)}_{K^-d}$ as follows
\begin{eqnarray}\label{labelA.1}
 M^{(1)}(K^-d \to K^-d) = 24\pi(m_K + m_d)
 a^{(1)}_{K^-d}(\vec{k}\,'\cdot \vec{k}\,),
\end{eqnarray}
where $\vec{k}$ and $\vec{k}\,'$ are momenta of a relative motion of
the $K^-d$ pair in the initial and final states. They are related by
$|\vec{k}\,| = |\vec{k}\,'|$. In turn, the P--wave amplitude
$M^{(1)}(K^-d \to K^-d)$ is defined in terms of the matrix element of
the $\mathbb{T}^{(1)}$--matrix as
\begin{eqnarray}\label{labelA.2}
 \langle K^-d|\mathbb{T}^{(1)}|K^-d\rangle  = (2\pi)^4\delta^{(4)}(k'_d
 + k' - k_d + k)\,M^{(1)}(K^-d \to K^-d),
\end{eqnarray}
where $(k_d, k)$, $(k'_d, k')$ are 4--momenta of the deuteron and
$K^-$--meson in the initial and final states, respectively. 

For the calculation of the matrix elements of the $\mathbb{T}$--matrix
we use the following effective Lagrangian
\begin{eqnarray}\label{labelA.3}
 {\cal L}_{\rm int}(x) = {\cal L}^{(0)}_{\rm int}(x) + {\cal
 L}^{(1)}_{\rm int}(x),
\end{eqnarray}
where the effective Lagrangians ${\cal L}^{(0)}_{\rm int}(x)$ and
 ${\cal L}^{(1)}_{\rm int}(x)$ define low--energy $K^-d$ interactions
in the S--wave and P--wave states. They are given by
\begin{eqnarray}\label{labelA.4}
\hspace{-0.3in}{\cal L}^{(0)}_{\rm int}(x) &=& 4\pi [\hat{a}^{(0)}_p
K^{-\dagger}(x) K^-(x) \bar{p}(x) p(x) + \hat{a}^{(0)}_x
\bar{K}^{0\dagger}(x) K^-(x) \bar{n}(x)p(x)] \nonumber\\
\hspace{-0.3in}&+& 4\pi [\hat{a}^{(0)}_n K^{-\dagger}(x)
K^-(x)\bar{n}(x)n(x) + \hat{a}^{(0)}_x K^{-\dagger}(x) \bar{K}^0(x)
\bar{p}(x) n(x)]\nonumber\\
\hspace{-0.3in}&+& 4\pi [\hat{a}^{0(0)}_n
\bar{K}^{0\dagger}(x)\bar{K}^0(x) \bar{n}(x)n(x)]
\end{eqnarray}
and 
\begin{eqnarray}\label{labelA.5}
\hspace{-0.3in}{\cal L}^{(1)}_{\rm int}(x) &=& 12\pi [\hat{a}^{(1)}_p
\bigtriangledown K^{-\dagger}(x) \cdot \bigtriangledown K^-(x)
\bar{p}(x) p(x) + \hat{a}^{(1)}_x
\bigtriangledown\bar{K}^{0\dagger}(x)\cdot \bigtriangledown K^-(x)
\bar{n}(x)p(x)] \nonumber\\
\hspace{-0.3in}&+& 12\pi [\hat{a}^{(1)}_n
\bigtriangledown K^{-\dagger}(x) \cdot
\bigtriangledown K^-(x)\bar{n}(x)n(x) + \hat{a}^{(1)}_x
\bigtriangledown K^{-\dagger}(x)\cdot \bigtriangledown \bar{K}^0(x)
\bar{p}(x) n(x)]\nonumber\\
\hspace{-0.3in}&+& 12\pi [\hat{a}^{0(1)}_n \bigtriangledown
 \bar{K}^{0\dagger}(x) \cdot \bigtriangledown \bar{K}^0(x) \bar{n}(x)
 n(x)].
\end{eqnarray}
The S--wave and P--wave scattering lengths are determined as shown in
Eq.(\ref{label3}). The matrix elements of these Lagrangians define the
S--wave and P--wave amplitudes of low--energy $\bar{K}N$ scattering,
expressed in terms of the S--wave and P--wave scattering lengths.

Using the effective Lagrangian Eq.(\ref{labelA.4}) one reproduces the
solution of the Faddeev equations in the fixed centre approximation
for the complex S--wave scattering length of $K^-d$ scattering, given
by Eq.(\ref{label2}) and obtained for the first time in \cite{Oset}.

For the calculation of the contributions of the single and double
scattering to the complex P--wave scattering length of $K^-d$
scattering we will use both Lagrangians Eq.(\ref{labelA.4}) and
Eq.(\ref{labelA.5}).  The $\mathbb{T}^{(1)}$--matrix for low--energy
$K^-d$ scattering, describing the contributions of the single and
double scattering, is defined by
\begin{eqnarray}\label{labelA.6}
\mathbb{T}^{(1)} = \int d^4x\,{\cal L}^{(0)}_{\rm int}(x) + i\int
d^4x_1d^4x_2\,{\rm T}({\cal L}^{(0)}_{\rm int}(x_1){\cal L}^{(1)}_{\rm
int}(x_2)) + \ldots,
\end{eqnarray}
where ${\rm T}$ is the time--ordering operator and the ellipsis
denotes the triple scattering contributions and so on. The P--wave
amplitude of low--energy $K^-d$ scattering, caused by the single and
double scattering, is
\begin{eqnarray}\label{labelA.7}
M^{(1)}(K^-d \to K^-d) = M^{(1)}(K^-d \to K^-d)_{\rm s.sc.} +
M^{(1)}(K^-d \to K^-d)_{\rm d.sc.},
\end{eqnarray}
where $M^{(1)}(K^-d \to K^-d)_{\rm s.sc.}$ and $M^{(1)}(K^-d \to
K^-d)_{\rm
d.sc.}$ are the amplitudes of the single and double scattering,
respectively. They are given by
\begin{eqnarray}\label{labelA.8}
M^{(1)}(K^-d \to K^-d)_{\rm s.sc.} &=& \langle K^-d|{\cal
L}^{(0)}_{\rm int}(0)|K^-d\rangle,\nonumber\\ M^{(1)}(K^-d \to
K^-d)_{\rm d.sc.} &=& i\int d^4x \langle K^-d|{\rm T}({\cal
L}^{(0)}_{\rm int}(x){\cal L}^{(1)}_{\rm int}(0))|K^-d\rangle.
\end{eqnarray}
For the calculation of the matrix element $\langle
K^-d|\mathbb{T}^{(1)}|K^-d\rangle$ we use the following the wave
functions of the initial and final states
\begin{eqnarray}\label{labelA.9}
|K^-d\rangle &=&
c^{\dagger}_{K^-}(\vec{k}\,)|d(-\vec{k},\lambda)\rangle,\nonumber\\
\langle K^-d| &=& \langle d(-\vec{k}\,',\lambda)|c_{K^-}(\vec{k}\,'),
\end{eqnarray}
where $\vec{k}$ and $\vec{k}\,'$ are the relative momenta of the
$K^-d$ pairs in the initial and final states, respectively,
$c^{\dagger}_{K^-}(\vec{k}\,)$ and $c_{K^-}(\vec{k}\,')$ are operators
of creation and annihilation of the $K^-$--mesons with 3--momenta
$\vec{k}$ and $\vec{k}\,'$, respectively. They obey standard
relativistic covariant commutation relations \cite{KDS6}. The wave
function of the deuteron $|d(-\vec{k},\lambda)\rangle$ is taken in the
momentum and particle number representation. It reads \cite{KDS6}
\begin{eqnarray}\label{labelA.10}
\hspace{-0.3in}|d(-\vec{k}, \lambda)\rangle = \frac{\sqrt{2
E_d(\vec{k}\,)}}{(2\pi)^3}\int \frac{d^3k_p}{\sqrt{2
E_N(\vec{k}_p)}}\frac{d^3k_n}{\sqrt{2
E_N(\vec{k}_n)}}\,\delta^{(3)}(\vec{k} + \vec{k}_p +
\vec{k}_n)\tilde{\Phi}_d\Big(\frac{\vec{k}_p - \vec{k}_n}{2}\Big)\,
a^{\dagger}_p(\vec{k}_p, \sigma_p)
a^{\dagger}_n(\vec{k}_n,\sigma_n)|0\rangle,
\end{eqnarray}
where $E_d(\vec{k}\,)$, $E_N(\vec{k}_p)$ and $E_N(\vec{k})$ are the
total energies of the deuteron, proton and neutron, respectively,
$\tilde{\Phi}_d(\vec{q}\,)$ is the wave function of the ground state
of the deuteron in the momentum representation,
$a^{\dagger}_p(\vec{k}_p,\sigma_p)$ and
$a^{\dagger}_n(\vec{k}_n,\sigma_n)$ are the operators of creation of
the proton and the neutron with 3--momenta $\vec{k}_p$ and $\vec{k}_n$
and polarisations $\sigma_p = \pm \frac{1}{2}$ and $\sigma_n = \pm
\frac{1}{2}$, respectively, and they obey standard relativistic
covariant anti--commutation relations \cite{KDS6}, $\lambda = \sigma_p
+ \sigma_n$ is the polarisation of the deuteron, and $|0\rangle$ is
the vacuum wave function. For the deuteron polarisation states with
$\lambda = \pm 1$ and $\lambda = 0$ the product
$a^{\dagger}_p(\vec{k}_p, \sigma_p) a^{\dagger}_n(\vec{k}_n,\sigma_n)$
should be replaced by $a^{\dagger}_p(\vec{k}_p, \pm \frac{1}{2})
a^{\dagger}_n(\vec{k}_n, \pm \frac{1}{2})$ and
$\frac{1}{\sqrt{2}}(a^{\dagger}_p(\vec{k}_p, \sigma_p)
a^{\dagger}_n(\vec{k}_n,- \sigma_p) + a^{\dagger}_p(- \vec{k}_p,
\sigma_p) a^{\dagger}_n(\vec{k}_n,\sigma_p))$, respectively
\cite{KDS6}.

The P--wave amplitude of the single $K^-d$ scattering is equal to
\begin{eqnarray}\label{labelA.11}
\hspace{-0.3in}&&M^{(1)}(K^-d \to K^-d)_{\rm s.sc.} = \langle
K^-d|{\cal L}^{(0)}_{\rm int}(0)|K^-d\rangle = 24\pi m_d
(\hat{a}^{(1)}_p + \hat{a}^{(1)}_n) (\vec{k}\,'\cdot \vec{k}\,)\int
\frac{d^3q}{(2\pi)^3}\,\tilde{\Phi}^*_d\Big(\vec{q} +
\frac{1}{2}\,\vec{k}\,'\Big)\tilde{\Phi}_d\Big(\vec{q} +
\frac{1}{2}\,\vec{k}\,\Big).\nonumber\\ \hspace{-0.3in}&&
\end{eqnarray}
The momentum integral defines the form factor $F_d(\vec{Q}\,)$ of the
deuteron \cite{Oset}
\begin{eqnarray}\label{labelA.12}
\hspace{-0.3in}&&\int
\frac{d^3q}{(2\pi)^3}\,\tilde{\Phi}^*_d\Big(\vec{q} +
\frac{1}{2}\,\vec{k}\,'\Big)\tilde{\Phi}_d\Big(\vec{q} +
\frac{1}{2}\,\vec{k}\,\Big) = \int
d^3x\,|\Phi_d(\vec{r}\,)|^2\,e^{\,i\vec{Q}\cdot \vec{r}} =
F_d(\vec{Q}\,),
\end{eqnarray}
where $\vec{Q} = \frac{1}{2}\,(\vec{k}\,' - \vec{k}\,)$ is the
momentum transfer. Since the form factor of the deuteron is normalised
to unity at $\vec{Q} = 0$ \cite{Oset}, the complex P--wave scattering
length, calculated in the single scattering approximation, is equal
to
\begin{eqnarray}\label{labelA.13}
(\tilde{a}^{(1)}_{K^-d})_{\rm s.sc.} = \frac{m_d}{m_K +
m_d}(\hat{a}^{(1)}_p + \hat{a}^{(1)}_n) = -\,0.231 + i\, 0.645\,{\rm fm^3},
\end{eqnarray}
where we have used the numerical values of the P--wave scattering
lengths of $K^-p$ and $K^-n$ scattering, adduced in Eq.(\ref{label13}).

The amplitude $M^{(1)}(K^-d \to K^-d)_{\rm d.sc.}$ of the double
scattering is defined by the matrix element
\begin{eqnarray}\label{labelA.14}
\hspace{-0.3in}&&M^{(1)}(K^-d \to K^-d)_{\rm d.sc.} = i \int
d^4x\,\langle K^-d|{\rm T}({\cal L}^{(0)}_{\rm int}(x){\cal
L}^{(1)}_{\rm int}(0)|K^-d\rangle = \nonumber\\
\hspace{-0.3in}&& = 48\pi^2\,i \int d^4x\langle
K^-d|\Big\{\hat{a}^{(0)}_p \hat{a}^{(1)}_n [\bar{p}(x)p(x)]{\rm
T}\Big(K^{-\dagger}(x)K^-(x) \nabla K^{-\dagger}(0)\cdot \nabla
K^-(0)\Big)[\bar{n}(0)n(0)]\nonumber\\
\hspace{-0.3in}&&\hspace{1.3in}+ \hat{a}^{(0)}_n
\hat{a}^{(1)}_p [\bar{n}(x)n(x)]{\rm T}\Big(K^{-\dagger}(x)K^-(x)
\nabla K^{-\dagger}(0)\cdot \nabla K^-(0)\Big)[\bar{p}(0)p(0)]\nonumber\\
\hspace{-0.3in}&&\hspace{1.3in}+ \hat{a}^{(0)}_x \hat{a}^{(1)}_x
[\bar{p}(x)n(x)]{\rm T}\Big(K^{-\dagger}(x)\bar{K}^0(x) \nabla
\bar{K}^{0\dagger}(0)\cdot \nabla K^-(0)\Big)[\bar{n}(0)p(0)]\nonumber\\
\hspace{-0.3in}&&\hspace{1.3in}+ \hat{a}^{(0)}_x \hat{a}^{(1)}_x
[\bar{n}(x)p(x)]{\rm T}\Big(\bar{K}^{0\dagger}(x)K^-(x) \nabla
K^{-\dagger}(0)\cdot \nabla
\bar{K}^0(0)\Big)[\bar{p}(0)n(0)]\Big\}|K^-d\rangle
\end{eqnarray}
Having calculated the matrix element between the $K^-$--meson states,
we arrive at the expression
\begin{eqnarray}\label{labelA.15}
\hspace{-0.3in}&&M^{(1)}(K^-d \to K^-d)_{\rm d.sc.} = 48\pi^2\,i \int
d^4x\Bigg\{\hat{a}^{(0)}_p
\hat{a}^{(1)}_n \langle d|[\bar{p}(x)p(x)][\bar{n}(0)n(0)]|d\rangle
\Big(\int \frac{d^4q}{(2\pi)^4 i}\,\frac{(\vec{k}\cdot \vec
{q}\,)\,e^{-i(q - k')\cdot x}}{m^2_K - q^2 -i0}\nonumber\\
\hspace{-0.3in}&& + \int \frac{d^4q}{(2\pi)^4 i}\,\frac{(\vec{k}'\cdot
\vec {q}\,)\,e^{+i(q - k)\cdot x}}{m^2_K - q^2 -i0}\Big) +
\hat{a}^{(0)}_n \hat{a}^{(1)}_p \langle
d|[\bar{n}(x)n(x)][\bar{p}(0)p(0)]|d\rangle \Big(\int
\frac{d^4q}{(2\pi)^4 i}\,\frac{(\vec{k}\cdot \vec{q}\,)\,e^{-i(q -
k')\cdot x}}{m^2_K - q^2 -i0}\nonumber\\
\hspace{-0.3in}&& + \int \frac{d^4q}{(2\pi)^4 i}\,\frac{(\vec{k}'\cdot
\vec {q}\,)\,e^{+i(q - k)\cdot x}}{m^2_K - q^2 -i0}\Big) +
\hat{a}^{(0)}_x \hat{a}^{(1)}_x \langle
d|[\bar{p}(x)n(x)][\bar{n}(0)p(0)]|d\rangle \int \frac{d^4q}{(2\pi)^4
i}\,\frac{(\vec{k}\cdot \vec {q}\,)\,e^{-i(q - k')\cdot x}}{m^2_K -
q^2 -i0}\nonumber\\
\hspace{-0.3in}&& + \hat{a}^{(0)}_x \hat{a}^{(1)}_x \langle
d|[\bar{n}(x)p(x)][\bar{p}(0)n(0)]|d\rangle \int \frac{d^4q}{(2\pi)^4
i}\,\frac{(\vec{k}'\cdot \vec {q}\,)\,e^{+i(q - k)\cdot x}}{m^2_K -
q^2 -i0}\Bigg\}.
\end{eqnarray}
The matrix elements of the products of the nucleon field operators
between the deuteron states, calculated in the non--relativistic
approximation, are equal to
\begin{eqnarray}\label{labelA.16}
\hspace{-0.3in}&&\langle d|[\bar{p}(x)p(x)][\bar{n}(0)n(0)]|d\rangle =
\langle d|[\bar{n}(x)n(x)][\bar{p}(0)p(0)]|d\rangle = - \langle
d|[\bar{p}(x)n(x)][\bar{n}(0)p(0)]|d\rangle = - \langle
d|[\bar{n}(x)p(x)][\bar{p}(0)n(0)]|d\rangle =\nonumber\\
\hspace{-0.3in}&&= 2 m_d \int
\frac{d^3q'}{(2\pi)^3}\frac{d^3q}{(2\pi)^3}\,\tilde{\Phi}^*_d\Big(\vec{q}\,'
+ \frac{1}{2}\,\vec{k}'\,\Big)\tilde{\Phi}_d\Big(\vec{q} +
\frac{1}{2}\,\vec{k}\,\Big)\,e^{\,-\,i(\vec{q}\,' - \vec{q}\,)\cdot
\vec{r}} = 2 m_d |\Phi_d(\vec{r}\,)|^2\,e^{\,i\frac{1}{2}(\vec{k}' -
\vec{k}\,)\cdot \vec{r}}.
\end{eqnarray}
Substituting Eq.(\ref{labelA.16}) into Eq.(\ref{labelA.15}) and
integrating over time we transcribe the r.h.s of Eq.(\ref{labelA.15})
into the form
\begin{eqnarray}\label{labelA.17}
\hspace{-0.3in}&&M^{(1)}(K^-d \to K^-d)_{\rm d.sc.} = 96\pi^2 m_d \int
d^3x\,|\Phi_d(\vec{r}\,)|^2\,e^{\,i\frac{1}{2}(\vec{k}' -
\vec{k}\,)\cdot \vec{r}}\Big(\hat{a}^{(0)}_p \hat{a}^{(1)}_n +
\hat{a}^{(0)}_n \hat{a}^{(1)}_p - \hat{a}^{(0)}_x
\hat{a}^{(1)}_x\Big)\nonumber\\
\hspace{-0.3in}&& \times \Big(\int \frac{d^4q}{(2\pi)^3 }\,\delta(q_0
- m_K)\frac{(\vec{k}\cdot \vec {q}\,)\,e^{\,+ i(\vec{q} -
    \vec{k}')\cdot \vec{r}}}{m^2_K - q^2 -i0}+ \int
\frac{d^4q}{(2\pi)^3}\,\delta(q_0 - m_K)\frac{(\vec{k}'\cdot \vec
  {q}\,)\,e^{-i(\vec{q} - \vec{k}\,)\cdot \vec{r}}}{m^2_K - q^2
  -i0}\Big),
\end{eqnarray}
where we have neglected the contributions of the kinetic energies of
the $K^-$--mesons with respect to their masses in the initial and
final states.  Having integrated over $q_0$ we obtain the r.h.s. of
Eq.(\ref{labelA.17}) in the following form
\begin{eqnarray}\label{labelA.18}
\hspace{-0.3in}&&M^{(1)}(K^-d \to K^-d)_{\rm d.sc.} = 96\pi^2 m_d \int
d^3x\,|\Phi_d(\vec{r}\,)|^2\,\Big(\hat{a}^{(0)}_p \hat{a}^{(1)}_n +
\hat{a}^{(0)}_n \hat{a}^{(1)}_p - \hat{a}^{(0)}_x
\hat{a}^{(1)}_x\Big)\nonumber\\
\hspace{-0.3in}&& \times \Big(e^{\,- i\frac{1}{2}(\vec{k}' +
  \vec{k}\,)\cdot \vec{r}}\int \frac{d^3q}{(2\pi)^3
}\,\frac{(\vec{k}\cdot \vec {q}\,)}{\vec {q}^{\;2}}\,e^{\,+
  i\vec{q}\cdot \vec{r}}+ e^{\,+i\frac{1}{2}(\vec{k}' +
  \vec{k}\,)\cdot \vec{r}}\int
\frac{d^3q}{(2\pi)^3}\,\frac{(\vec{k}'\cdot \vec
  {q}\,)}{\vec{q}^{\;2}}\,e^{\,-i \vec{q}\cdot \vec{r}}\Big)
=\nonumber\\
\hspace{-0.3in}&& = 96\pi^2 m_d \int
d^3x\,|\Phi_d(\vec{r}\,)|^2\,\Big(\hat{a}^{(0)}_p \hat{a}^{(1)}_n +
\hat{a}^{(0)}_n \hat{a}^{(1)}_p - \hat{a}^{(0)}_x
\hat{a}^{(1)}_x\Big)\,e^{\,- i\frac{1}{2}(\vec{k}' + \vec{k}\,)\cdot
\vec{r}}\,(\vec{k}' + \vec{k}\,)\cdot (-i\nabla)\int \frac{d^3q}{(2\pi)^3
}\,\frac{e^{\,+ i\vec{q}\cdot \vec{r}}}{\vec {q}^{\;2}} = \nonumber\\
\hspace{-0.3in}&& = 24\pi m_d \int
d^3x\,|\Phi_d(\vec{r}\,)|^2\,\Big(\hat{a}^{(0)}_p \hat{a}^{(1)}_n +
\hat{a}^{(0)}_n \hat{a}^{(1)}_p - \hat{a}^{(0)}_x
\hat{a}^{(1)}_x\Big)\,e^{\,- i\frac{1}{2}(\vec{k}' + \vec{k}\,)\cdot
\vec{r}}\,\frac{i(\vec{k}' + \vec{k}\,)\cdot \vec{r}}{r^3},
\end{eqnarray}
where we have made a change of variables $\vec{r} \to -\,\vec{r}$
in the second term and have used that $|\Phi_d(-\,\vec{r}\,)|^2 =
|\Phi_d(\vec{r}\,)|^2$.

Expending the exponential $e^{\,- i\frac{1}{2}(\vec{k}' +
  \vec{k}\,)\cdot \vec{r}}$ in powers of $\frac{1}{2}(\vec{k}' +
  \vec{k}\,)\cdot \vec{r}$, we obtain the contribution of the double
  scattering to P--wave amplitude of $K^-d$ scattering in the form
\begin{eqnarray}\label{labelA.19}
\hspace{-0.3in}M^{(1)}(K^-d \to K^-d)_{\rm d.sc.} = 24\pi m_d\,
\frac{1}{3}\,\Big(\hat{a}^{(0)}_p \hat{a}^{(1)}_n + \hat{a}^{(0)}_n
\hat{a}^{(1)}_p - \hat{a}^{(0)}_x \hat{a}^{(1)}_x\Big)(\vec{k}'\cdot
\vec{k}\,)\int \frac{d^3x}{r}\,|\Phi_d(\vec{r}\,)|^2,
\end{eqnarray}
 where we have omitted the terms proportional to $|\vec{k}'|^2$ and
 $|\vec{k}\,|^2$, which have no relation to $K^-d$ scattering in the
 P--wave state.  The terms may contribute to the S--wave amplitude of
 $K^-d$ scattering, defining the effective range of $K^-d$ scattering
 in the S--wave state, but vanish in the complex S--wave scattering
 length of $K^-d$ scattering, calculated at $\vec{k}, \vec{k}\,' \to
 0$.

The contribution of the double scattering to the complex P--wave
scattering length of $K^-d$ scattering is
\begin{eqnarray}\label{labelA.20}
\hspace{-0.3in}(\tilde{a}^{(1)}_{K^-d})_{\rm d.sc.} = \frac{m_d}{m_K
+ m_d}\, \frac{1}{3}\,\Big(\hat{a}^{(0)}_p \hat{a}^{(1)}_n +
\hat{a}^{(0)}_n \hat{a}^{(1)}_p - \hat{a}^{(0)}_x
\hat{a}^{(1)}_x\Big)\int \frac{d^3x}{r}\,|\Phi_d(\vec{r}\,)|^2 =
-\,0.031 -i\,0.097\,{\rm fm^3},
\end{eqnarray}
where we have used the numerical values of the S--wave and P--wave
scattering lengths of $\bar{K}N$ scattering, adduced in
Eq.(\ref{label13}). Thus, the complex P--wave scattering length of
$K^-d$ scattering, defined by the contributions of the single and
double scattering, is equal to
\begin{eqnarray}\label{labelA.21}
\hspace{-0.3in}\tilde{a}^{(1)}_{K^-d} = \frac{m_d}{m_K +
m_d}\,\Big(\hat{a}^{(1)}_p + \hat{a}^{(1)}_n +
\frac{1}{3}\,(\hat{a}^{(0)}_p \hat{a}^{(1)}_n + \hat{a}^{(0)}_n
\hat{a}^{(1)}_p - \hat{a}^{(0)}_x \hat{a}^{(1)}_x)\int
\frac{d^3x}{r}\,|\Phi_d(\vec{r}\,)|^2\Big) = -\,0.262 +
i\,0.548\,{\rm fm^3}.
\end{eqnarray}
The contribution of the triple scattering we obtain by using the exact
solution. It reads
\begin{eqnarray}\label{labelA.22}
\hspace{-0.3in}&&(\tilde{a}^{(1)}_{K^-d})_{\rm tr.sc.} =
\frac{m_d}{m_K + m_d}\,\frac{1}{36}\Big[\hat{a}^{(1)}_p\Big(7
\hat{a}^{(0)}_p \hat{a}^{(0)}_n + (\hat{a}^{(0)}_n)^2 -
(\hat{a}^{0(0)}_n)^2\Big) + \hat{a}^{(1)}_n\Big(7 (\hat{a}^{(0)}_p
\hat{a}^{(0)}_n - (\hat{a}^{(0)}_x)^2) + \hat{a}^{(0)}_p (\hat{a}^{(0)}_n +
\hat{a}^{0(0)}_n)\nonumber\\
\hspace{-0.3in}&& - 2 \hat{a}^{(0)}_n \hat{a}^{(0)}_x\Big)+
\hat{a}^{(1)}_x \hat{a}^{(0)}_x\Big(\hat{a}^{0(0)}_n -
\hat{a}^{(0)}_n\Big) + \hat{a}^{0(1)}_n(\hat{a}^{(0)}_x)^2\Big]\int
\frac{d^3x}{r^2}\,|\Phi_d(\vec{r}\,)|^2 = -\,0.015 -\,i\,0.023\,{\rm fm^3}.
\end{eqnarray}
The complex P--wave scattering length of $K^-d$ scattering, accounting
for the contributions of the single, double and triple scattering, is
equal to $\tilde{a}^{(1)}_{K^-d} = (\tilde{a}^{(1)}_{K^-d})_{\rm
s.sc.} + (\tilde{a}^{(1)}_{K^-d})_{\rm d.sc.} +
(\tilde{a}^{(1)}_{K^-d})_{\rm tr.sc.} = - 0.277\, + i\, 0.525\,{\rm
fm^3}$. The discrepancy of this value with the complex P--wave
scattering length, defined by the solution of the Faddeev equations
Eq.(\ref{label8}), is $\delta \tilde{a}^{(1)}_{K^-d} = - 0.075
-i\,0.072\,{\rm fm^3}$. It makes up of about $21\,\%$ and $17\,\%$ of
the real and imaginary parts of the total complex P--wave scattering
length, respectively.

\subsection{Appendix B: Complex S--wave and P--wave scattering length of
 $\bar{K}N$ scattering} \renewcommand{\theequation}{B-\arabic{equation}}
\setcounter{equation}{0}

In this Appendix we outline our procedure for the calculation of the
complex S--wave and P--wave scattering lengths of $\bar{K}N$
scattering. Following \cite{ECL1}, the amplitude $M_0(\bar{K}N \to
PB)$ for the $\bar{K}N \to PB$ reaction, where $P$ is a pseudoscalar
meson and $B$ is a ground--state baryon, we calculate in the
tree--approximation. For this aim we use the chiral Lagrangian
Eq.(\ref{label14}) and the chiral Lagrangians
\begin{eqnarray}\label{labelB.1}
\hspace{-0.3in}{\cal L}_{\rm int}[\Lambda^*(x),B(x),P(x)] &=&
    g_{\Lambda^*}\,\bar{\Lambda}^*(x)\gamma^{\mu}\gamma^5 \langle
    p_{\mu}(x)B(x)\rangle,\nonumber\\
\hspace{-0.3in}{\cal L}_{\rm int}[B_j(x),B(x),P(x)] &=&\langle
  \bar{B}_j(x)i \gamma^{\mu}[s_{\mu}(x),B(x)]\rangle - g_{A_j}\,(1 -
  \alpha_{D_j})\,
  \langle\bar{B}_j(x)\gamma^{\mu}[p_{\mu}(x),B(x)]\rangle\nonumber\\
\hspace{-0.3in}&-& \,g_{A_j}\,\alpha_{D_j}\, \langle
  \bar{B}_j(x)\gamma^{\mu}\{p_{\mu}(x),B(x)\}\rangle,
\end{eqnarray}
describing the low--energy interactions invariant under chiral
$SU(3)\times SU(3)$ symmetry of the $\Lambda(1405)$ resonance,
defined by the field operator $\Lambda^*(x)$, and the baryon
resonances $B_j(\underline{\bf 8})$, defined by the field operators
$B_j(x)$ for $j = 1,2$ \cite{IV5}, with the ground--state baryon octet
$B(x)$ and the octet $P(x)$ of pseudoscalar mesons. 

In addition we take into account the interactions, invariant under
chiral $SU(3)\times SU(3)$ symmetry, of the $\Delta(1232)$ resonance
and $\Sigma(1385)$ resonance, defined by the field operators
$D^{abc}_{\mu}(x)$, with the ground--state baryon octet $B(x)$ and the
octet $P(x)$ of pseudoscalar mesons. They are defined by the chiral
Lagrangian
\begin{eqnarray}\label{labelB.2}
  \hspace {-0.3in}{\cal L}_{\rm int}[D(x), B(x),P(x)] = \sqrt{2}\,
g_{\Delta} \bar{D}^{abc}_{\mu}(x)\Theta^{\mu\nu}\gamma^5
(p_{\nu}(x))^d_a B^e_b(x)\,\varepsilon_{cde} + {\rm h.c.},
\end{eqnarray}
where the tensor $\Theta^{\mu\nu}$ is given in \cite{Delta3}:
$\Theta^{\mu\nu} = g^{\mu\nu} - (Z + 1/2)\gamma^{\mu}\gamma^{\nu}$,
where the parameter $Z$ is arbitrary.  There is no consensus on the
exact value of $Z$. From theoretical point of view $Z=1/2$ is
preferred \cite{Delta3}.  Phenomenological studies give only the bound
$|Z| \le 1/2$ \cite{Delta7,Delta8}. For the components of the decuplet
$D_{abc}(x)$ we use the following definition
\begin{eqnarray}\label{labelB.3}
  D_{111}(x) &=& \Delta^{++}(x),\; D_{112}(x) = \frac{1}{\sqrt{3}}\,
\Delta^+(x),\; D_{122}(x) = \frac{1}{\sqrt{3}}\,\Delta^0(x) ,\;
D_{222}(x) = \Delta^-(x),\nonumber\\ D_{113}(x) &=&
\frac{1}{\sqrt{3}}\,\Sigma^{*+}(x),\; D_{123}(x) =
\frac{1}{\sqrt{6}}\, \Sigma^{*0}(x),\; D_{223}(x) =
\frac{1}{\sqrt{3}}\,\Sigma^{*-}(x),\nonumber\\ D_{133}(x) &=&
\frac{1}{\sqrt{3}}\,\Xi^{*0}(x),\; D_{233}(x) =
\frac{1}{\sqrt{3}}\,\Xi^{*-}(x),\nonumber\\ D_{333}(x) &=&\Omega^-(x).
\end{eqnarray}
According to \cite{PDG10}, baryon resonances $B_1(\underline{\bf 8}) =
(N(1440), \Lambda(1600), \Sigma(1660))$ and $B_2(\underline{\bf 8}) =
(N(1710), \Lambda(1810), \Sigma(1880))$ belong to octets of $SU(3)_f$
symmetry \cite{PDG10} with the coupling constants $g_{A_1} = 0.62$,
$\alpha_{D_1} = 0.85$ and $g_{A_2} = 0.12$, $\alpha_{D_2} =
-\,1.55$. The experimental value of $g_{\Delta}$ is $g^{\exp}_{\Delta}
= (1.11\pm 0.04)\,g_A$ \cite{Delta4}, where $g_A = 1.2750$
\cite{Abele1,Faber3}.  The coupling constant of the
$\Lambda(1405)$ resonance we take equal to $g_{\Lambda^*} =
0.504$. It defines the width $\Gamma_{\Lambda^*} = 40\,{\rm MeV}$,
which fits well the imaginary part of the complex S--wave scattering
length of $K^-p$ scattering, measured recently by the SIDDHARTA
Collaboration.

The scalar resonances $f_0(980)$ and $a_0(980)$ with quantum numbers
$I(J^P) = 0(0^+)$ and $I(J^P) = 1(0^+)$ \cite{PDG10}, respectively,
give contributions to the $t$--channels of elastic and inelastic
$\bar{K}N$ scattering. According to Jaffe \cite{Jaffe}, the scalar
mesons $f_0(980)$ and $a_0(980)$ are four--quark states (or $\bar{K}K$
molecule), which belong to an $SU(3)_{\rm flavour}$ nonet. According
Ecker {\it et al.} \cite{Ecker}, the interaction of the scalar meson
resonances with octets of pseudoscalar mesons with derivative
couplings invariant under chiral $SU(3)\times SU(3)$ symmetry takes
the form
\begin{eqnarray}\label{labelB.4}
  {\cal L}_S(x) = 2\sqrt{2}\,g_S\,{\rm
  tr}\{S(x)\partial_{\mu}U^{\dagger}\partial^{\mu}U(x)\}.
\end{eqnarray}
Here $S(x)$ is a nonet of scalar $qq\bar{q}\bar{q}$ mesons, defined by
\cite{Jaffe}
\begin{eqnarray}\label{labelB.5}
  S^b_a = \left(\begin{array}{llcl} {\displaystyle
        \frac{a^0_0}{\sqrt{2}} - \frac{\varepsilon}{2}} &
      \hspace{0.3in}a^+_0 & \kappa^+ \\ \hspace{0.3in}a^-_0
        &{\displaystyle - \frac{a^0_0}{\sqrt{2}} -
        \frac{\varepsilon}{2}} & \kappa^0 \\
      \hspace{0.15in}  \kappa^- & \hspace{0.3in}\bar{\kappa}^0 &
      {\displaystyle - \frac{f_0}{\sqrt{2}} + \frac{\varepsilon}{2}} \\
    \end{array}\right),
\end{eqnarray}
where $g_S$ is a phenomenological coupling constant. The components of
the nonet Eq.(\ref{labelB.5}) have the following quark structures:
$\vec{a}_0 = (a^+_0, a^0_0, a^-_0) = (s\bar{s}u\bar{d},
s\bar{s}(u\bar{u} - d\bar{d})/\sqrt{2}, d\bar{u}s\bar{s})$ is the
isotriplet of $a_0(980)$ mesons, $\kappa = (\kappa^+,\kappa^0) =
(u\bar{s}d\bar{d},d\bar{s}u\bar{u})$ and $\bar{\kappa} =
(\bar{\kappa}^0, - \kappa^-) = (s\bar{d}u\bar{u}, - s\bar{u}d\bar{d})$
are doublets of strange scalar four--quark states, $f_0 =
s\bar{s}(u\bar{u} - d\bar{d})/\sqrt{2}$ is the $f_0(980)$ meson and
$\varepsilon$ is the isoscalar scalar $\varepsilon(700)$ meson with
quark structure $\varepsilon = u\bar{d}d\bar{u}$ and mass
$m_{\varepsilon} = 700\,{\rm MeV}$ \cite{Jaffe}. The nonet $S(x)$ is
constructed in such a way that the $f_0(980)$ meson decouples from the
$\pi \pi$ states, whereas the $\varepsilon(700)$ meson couples to the
$\pi \pi$ states but decouples from the $\bar{K}K$ states
\cite{Jaffe}. This implies that the $\varepsilon(700)$ meson does not
contribute to the amplitude of $K^-p$ scattering. The value of the
coupling constant $g_S$ one can define from the experimental values of
the width of the $a_0 \pi\eta$ decay $\Gamma^{\exp}_{a_0} = (50 \div
100)\,{\rm MeV}$ and the $\kappa \to K\pi$ decay $\Gamma_{\kappa} =
(290\pm 21)\,{\rm MeV}$, if we identify the scalar meson resonance
$\kappa$ with the scalar meson $K^*_0(1430)$ having mass $m_{K^*_0} =
(1414 \pm 6)\,{\rm MeV}$ \cite{PDG10}. We get $g_S = 28\,{\rm MeV}$
that gives the width of the $a_0 \pi\eta$ decay equal to $\Gamma_{a_0}
= 59\,{\rm MeV}$. This agrees well with the experimental data
$\Gamma^{\exp}_{a_0} = (50 \div 100)\,{\rm MeV}$ \cite{PDG10}.

The interaction of the scalar meson resonances $S(x)$ with
ground--state baryon octets we define as
\begin{eqnarray}\label{labelB.6}
  {\cal L}_{SBB}(x) = g_D\,{\rm tr}\{\bar{B}(x)\{B(x),S(x)\}\} +
  g_F\,{\rm tr}\{\bar{B}(x)[B(x),S(x)]\},
\end{eqnarray}
where $g_D$ and $g_F$ are the phenomenological coupling constants of
the symmetric and antisymmetric $SBB$ interactions. The coupling
constant $g_F$ should be set zero $g_F = 0$, since the
$\varepsilon(700)$ meson does not couple to the $\bar{N}N$ pair
\cite{Jaffe}.

The amplitude of the $\bar{K}N \to PB$ reaction, calculated in the
tree--approximation with the Lagrangians Eq.(\ref{label14}),
Eq.(\ref{labelB.1}), Eq.(\ref{labelB.2}), Eq.(\ref{labelB.3}),
Eq.(\ref{labelB.4}) and Eq.(\ref{labelB.6}), takes the form
\begin{eqnarray}\label{labelB.7}
\hspace{-0.3in}&&M_0(\bar{K}N\to PB) = M^{(c)}(\bar{K}N\to PB) +
M^{(b)}(\bar{K}N\to PB) + M^{(\rm WT)}(\bar{K}N\to PB) +
M^{(s)}_{\Lambda^* }(\bar{K}N\to PB)\nonumber\\\hspace{-0.3in} &&+
M^{(s)}_{J^P = \frac{1}{2}^{\!+}}(\bar{K}N\to PB) + M^{(u)}_{J^P =
\frac{1}{2}^{\!+}}(\bar{K}N\to PB) + M^{(s)}_{J^P =
\frac{3}{2}^{\!+}}(\bar{K}N\to PB) + M^{(u)}_{J^P =
\frac{3}{2}^{\!+}}(\bar{K}N\to PB)\nonumber\\
\hspace{-0.3in}&&+ M^{(t)}_{J^P = 0^+}(\bar{K}N\to PB),
\end{eqnarray}
where the first two amplitudes are defined by the interactions with
the coupling constants $d_j$ for $j = 1,2,3,4$ and $b_{\ell}$ for
$\ell = 0,D,F$, respectively, the third amplitude is caused by the
Weinberg--Tomozawa interactions, the other amplitudes are defined by
the exchange of the $\Lambda(1405)$ resonance, the ground--state
baryons, the baryon resonances $B_j(\underline{\bf 8})$ for $j = 1,2$,
the $\Sigma(1385)$ and $\Delta(1232)$ resonances and the scalar meson
resonances, respectively, in the $s$--, $u$-- and $t$--channels.

Expanding the amplitudes Eq.(\ref{labelB.6}) in powers of relative
momenta $\vec{k}$ and $\vec{k}'$ and keeping only the terms
independent of relative momenta and proportional to the scalar product
$\vec{k}'\cdot \vec{k}$ we define the contributions to the complex
S--wave and P--wave scattering lengths of $\bar{K}N$ scattering. Using
the matrix equation $M^{- 1} = M^{-1}_0 - G$ we obtain the unitarised
amplitudes of the reactions $\bar{K}N \to PB$ in terms of the complex
S--wave scattering lengths of all scattering channels $\bar{K}N \to
PB$ for $PB = \bar{K}N$ and $\pi Y$, where $Y = \Sigma,\Lambda^0$
hyperons. The input parameters of the approach $d_j$ for $j = 1,2,3,4$
and $g_D$ are fitted from the experimental data on the complex S--wave
scattering length of the SIDDHARTA Collaboration, the ratios of the
cross sections of inelastic $K^-p$ scattering in the S--wave state,
measured at threshold of the $K^-p \to \pi Y$ reactions
\cite{Kpt1,Kpt2}, and the experimental cross sections of elastic and
inelastic $K^-p$ scattering \cite{Kp1}--\cite{Kp5}. The cross sections
for elastic and inelastic $K^-p$ scattering are calculated at the
account for the pure Coulomb scattering and the Coulomb interactions
for the pairs of charged particles in the initial and final states. As
a result of this fit we get the following numerical values of the
input parameters $d_1 = -\,0.389\,{\rm fm}$, $d_2 = -\,0.709\,{\rm
fm}$, $d_3 = +\,2.816\,{\rm fm}$ and $d_4 = -\,0.619\,{\rm fm}$. As
has been found, the contribution of the scalar meson resonances is not
essential for reasonable values of the coupling constant $g_D$. The
calculated complex S--wave scattering lengths of $\bar{K}N$ scattering
describe reasonably well the experimental data on elastic and
inelastic $K^-p$ scattering in the low--energy region
\cite{Kp1}--\cite{Kp5} not far above the threshold of the production
of the $\bar{K}^0n$ pair, which is equal to $k_0 \simeq 58\,{\rm MeV}$
in the centre of mass frame. Using the numerical values of the input
parameters $d_1 = -\,0.389\,{\rm fm}$, $d_2 = -\,0.709\,{\rm fm}$,
$d_3 = +\,2.816\,{\rm fm}$ and $d_4 = -\,0.619\,{\rm fm}$ we evaluate
the complex S--wave scattering lengths and, correspondingly, the
complex P--wave scattering lengths of $\bar{K}N$ scattering. The
imaginary parts of the S--wave and P--wave scattering lengths are
defined by the dominant contributions of the $\Lambda(1405)$ and
$\Sigma(1385)$ resonances, that agrees well with the analysis of
low--energy $\bar{K}N$ interactions in the S--wave and P--wave states,
applied to the problem of antikaon--nuclear quasibound states
\cite{Weise1}.

\end{document}